# Electronic structure and oxygen reduction on tunable $[Ti(IV)Pc]^{2+}$ and Ti(II)Pc titanyl-phthalocyanines: A quantum chemical prediction

Jeffrey Roshan De Lile[a,b], Thomas Heine[b], Su Zhou[a*]

Jeffrey_delile@yahoo.com, suzhou@tongji.edu.cn, thomas.heine@uni-leipzig.de

[a]School of Automotive Studies, Tongji University,

4800 Cao'an Road, 201804, Shanghai,

P.R. China

[b]Wilhelm-Ostwald-Institut für Physikalische und Theoretische Chemie,

Universität Leipzig,

Linnéstraße 2, 04103 Leipzig,

Germany

**Abstract**

The high cost of platinum-based catalysts has hampered the commercialization of polymer electrolyte membrane fuel cells (PEMFCs). Hence, the electronic structure and oxygen reduction ability of $[Ti(IV)Pc]^{2+}$, Ti(II)Pc titanyl-phthalocyanines, and their tailored peripheral and axial ligand complexes were theoretically investigated to determine non-precious cathode catalysts. Our results revealed that the peripherally substituted and unsubstituted Ti(II)Pc triplet complexes can spontaneously reduce peroxide. The singlet $[Ti(IV)Pc]^{2+}$ parent complex has a 6.45 eV barrier. However, fluorine substitution at peripheral positions reduced the energy barrier up to 0.45 eV. In addition, chlorine substitution has shown spontaneous peroxide reduction as in the case of triplets. The high catalytic activity of Ti(II)Pc complexes and singlet chlorine substituted complex is attributed to the optimal charge transfer between dioxygen molecule and the novel catalyst complexes. As a result, Ti(II)Pcs and chlorine substituted singlet complexes are considered as potential substitutions for the noble Pt-based catalyst for the PEMFCs.

**1. Introduction**

Polymer electrolyte membrane fuel cells (PEMFCs) have long been considered as promising energy conversion devices for automobile transportation applications due to their high energy conversion efficiency and environmental friendliness. The state-of-the-art electrocatalysts of the PEMFC anode and cathode electrodes are predominantly carbon-supported platinum-based nanoparticles. However, the platinum-based catalysts suffer from sluggish cathode kinetics, which limits the oxygen cathode performance, and high costs. Thus, research efforts to find alternative solutions have intensified. In recent years, more attention has been given to the development of non-noble metal catalysts to replacement of platinum (Pt). Moreover, many different non-noble metal catalysts have been thoroughly investigated for their utility in the oxygen reduction reaction (ORR). Some of these non-noble metal catalysts are metal-carbides [1] metal-nitrides [2], spinel [3,4] and perovskite-type [5,6] metal oxides, metal-oxynitrides [7-10], metal-carbonitrides [11,12], metal-chalcogenides [13], conductive polymers [14], enzymatic compounds [15-17], and $N_4$-chelate macrocyclic compounds [18-20]. However, the fundamental drawbacks of the proposed catalysts present challenges to their implementation as PEMFC catalysts. Metal-carbides, for instance, are easily oxidized to form $CO_2$ due to the highly oxidizing cathode environment in fuel cells [21]. Metal-carbonitrides produced higher amounts of peroxide ions in acidic media, which can significantly degrade fuel cell performance. Some of the metal-chalcogenides tend to deintercalate at positive electrode potentials. Enzymatic-compounds are highly sensitive to system temperature and pH, and its partial adsorption to the electrode can provide a greater challenge in fuel cell conditions [22]. Macrocyclic compounds suffer from instability issues in acidic conditions. However, recently published literature [23] provides a better understanding of N4-chelate compounds and providing solutions to stabilize the system in fuel cell environments.

Among the macrocyclic complexes, metallophthalocyanines (MPcs) are the most versatile materials due to their low preparation cost, low toxicity, and high chemical and thermal stability [24]. They are used in various applications, ranging from color pigments, catalysts, liquid crystals, photosensitizers, electrochromic substances, and optical data storage to cancer therapy [25]. It is accepted that the breaking of the O-O bond during the $H_2O_2$ formation process on the MPcs is the main reason for the 4e- pathway of the ORR. Iron-phthalocyanine-based catalysts are the best oxygen reduction catalysts available, due to their ability to perform a 4e- oxygen reduction. However, they suffer from stability issues under PEMFC cathode environment. On the other hand, Co-phthalocyanine (CoPc) reduces oxygen through 2e- pathway due to its inability of breaking O-O bond of $H_2O_2$ and possesses higher stability under PEMFC conditions [26]. Thus, the ideal MPc for PEMFC should have catalytic activity comparable to FePc and stability comparable to CoPc [27].

Most of the MPc complexes reported in the literature for ORR have M(II)Pc core, such as $Fe^{II}$, $Co^{II}$, $Ni^{II}$, $Cu^{II}$ and $Zn^{II}$ divalent ion complexes [22,23,28,29]. The higher oxidized metal centers, such as Ti, V, Mn, and Cr, are seldom studied for their electronic structure and ORR ability. Cook et al. [30] and Liao et al. [28] systematically investigated the electronic structure of 3d MPcs from Mn to Zn using X-ray absorption spectroscopy and density functional theory (DFT). They suggested that MnPc is a suitable alternative for the ORR in PEMFC due to its stability and catalytic activity that are comparable to CoPc and FePc, respectively. Liu et al. [31] studied a series of MPcs, including CrPc, and predicted that CrPc does not actively participate in ORR. Nonetheless, vanadium (VPc) [32] and titanium phthalocyanines (TiPc) were investigated, mainly in their highest oxidation states as singlet molecules [33-37]. However, the knowledge of titanylphthalocyanines (TiPcs) is still limited to oxo(phthalocyaninato)titanium (TiOPc), dichloro(phthalocyaninato)titanium (TiCl$_2$Pc) [38], axially complex ligand substituted species, and their derivatives [39].

To the best of our knowledge, there are no accounts of non-axial complexes of TiPcs and their ORR activity. Thus, in this article [Ti(IV)Pc]$^{2+}$ (singlet), Ti(II)Pc (triplet), their tailor-made peripherally substituted complexes and their substitution effects on ORR are investigated using DFT calculations with a water implicit solvation model.

## 2. Methods

The DFT calculations were carried out using the Amsterdam Density Functional (ADF) software [40]. The exchange-correlation energies are described by the Perdew-Burke-Ernzerhof functional (PBE), a generalized gradient approximation (GGA) level of theory [41]. This functional is widely used in the literature, providing more availability of systems for comparison. In addition, some [39] found that X-ray absorption spectra are better reproduced with the PBE functional for TiOPc. The PBE functional with Grimme correction (PBE-D3) [41,42] was utilized to study all the molecules to account for weak interaction forces. Spin polarized calculations have been used to account for the correct multiplicity of the systems. All electron calculations have been employed in the present study to circumvent the errors that may occur due to the pseudopotential approximation. The Slater-type orbitals triple zeta plus two polarization function (TZ2P) basis set has been used in the calculations. Scalar relativistic effects account for contraction, decontraction and electronic stability of the orbitals; therefore, the ZORA scalar relativistic method [43] was used for all of the molecular optimizations in this study. To analyze water solvation effects, an implicit solvation method known as the conductor-like screening model (COSMO) [44] was used.

All of the structure optimizations were conducted using Becke normal integration accuracy with a spline fit. All of the molecules were fully relaxed until the maximum force was below 0.04 eV/Å to obtain the ground state energies. The Atoms in Molecule (AIM) [45] atomic charge population analysis was conducted to study the charge transfer effects in each of the molecular complexes. The linear transit (LT) and transition state search [46,47] methods implemented in the ADF program were employed to search for the transition state structure and activation energy for $H_2O_2$ dissociation into hydroxyl (OH). The LT oxygen-oxygen bond length of $H_2O_2$ was used as a constraint. Then, the constraint optimization was performed until the O-O bond dissociates to produce hydroxyl groups. All the high energy structures were optimized with Becke good integration accuracy, and single point calculations have been done to obtain vibrational frequencies of the optimized structures. The structure with one imaginary frequency was chosen as the transition state.

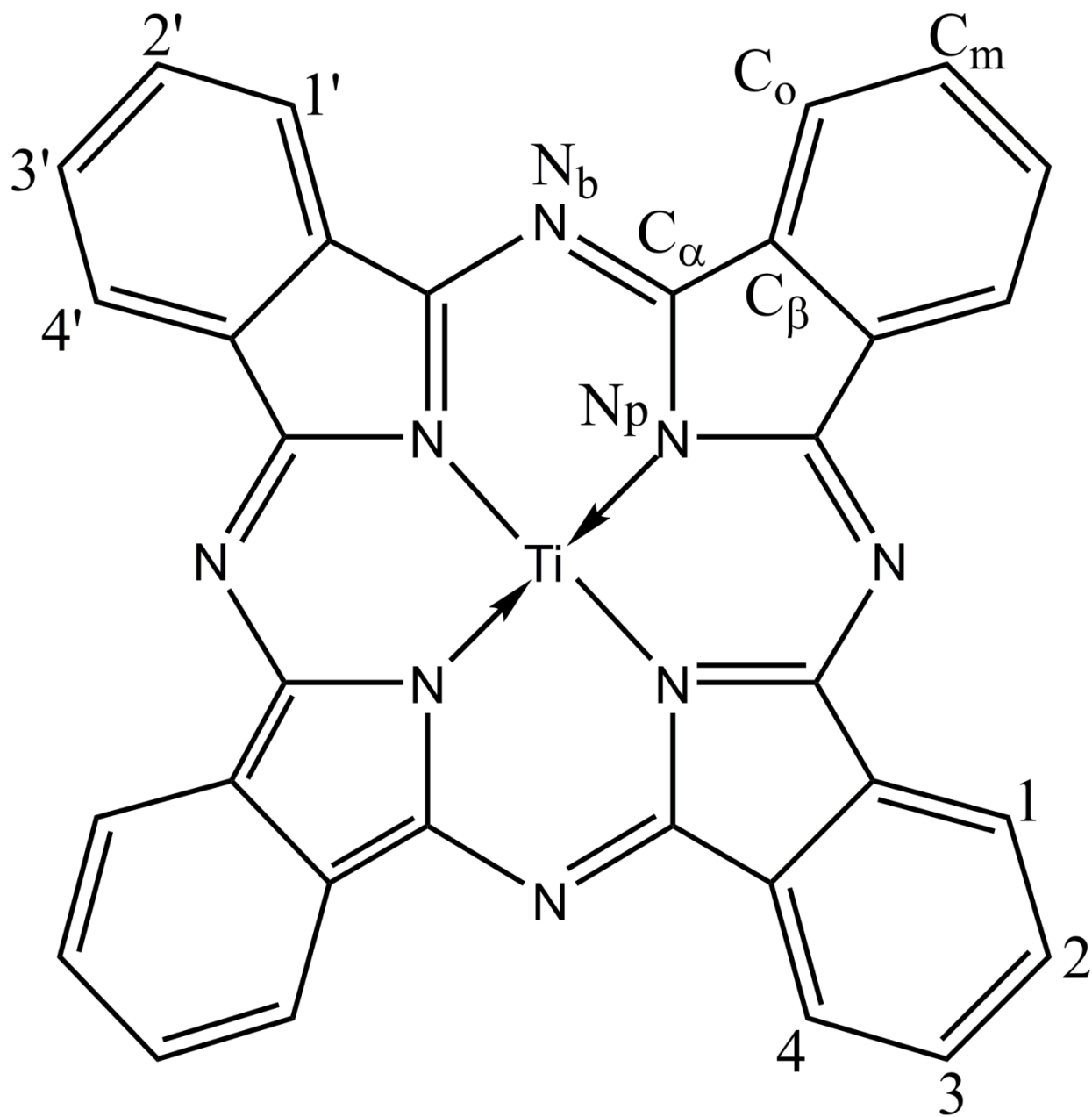

**Figure 1** The orientation and atom labeling scheme for TiPc molecules.

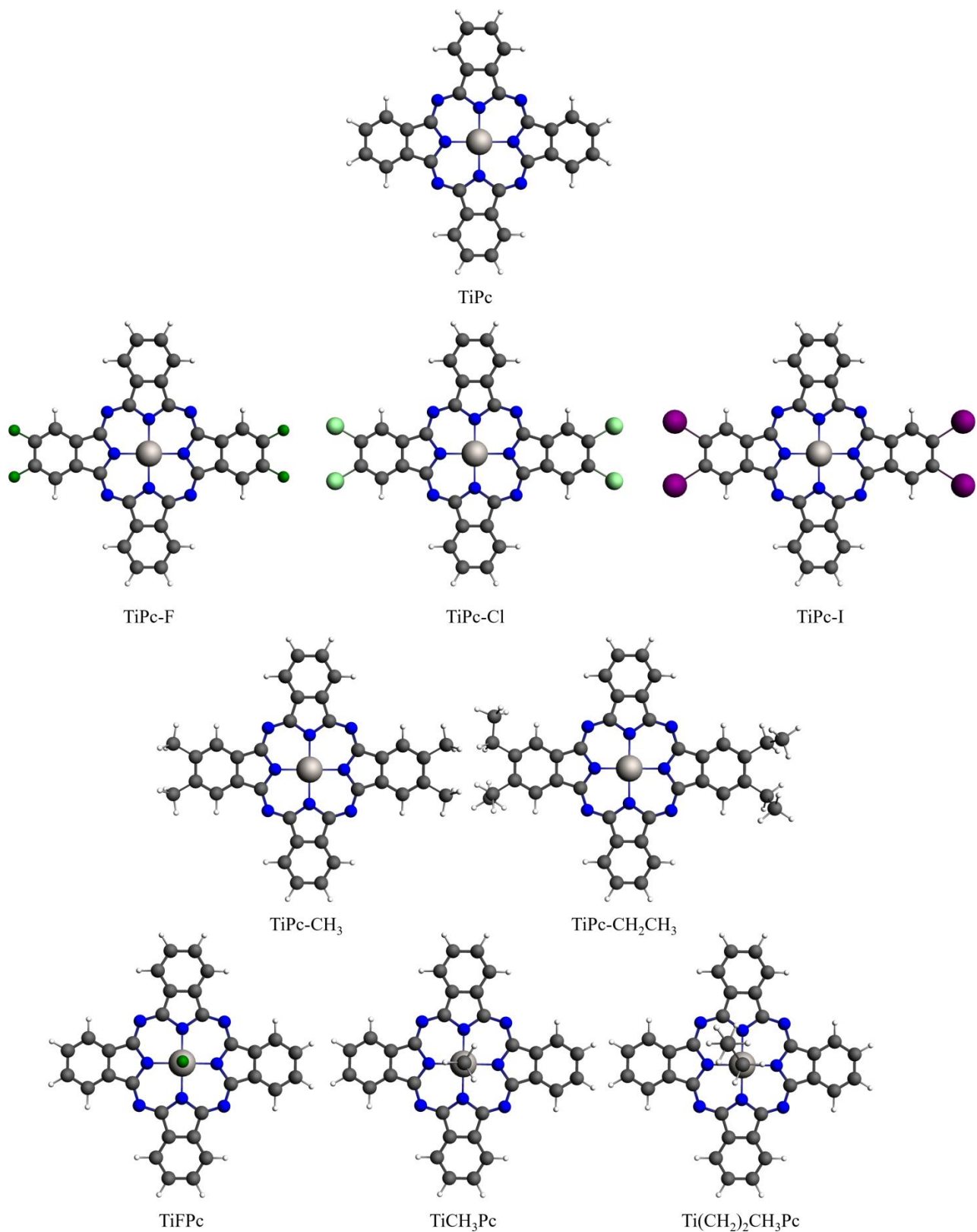

**Figure 2** Model molecular structures and their abbreviations. Peripherally halogen and methyl substituted complexes are D$_{2h}$ in point group symmetry. Ethyl substituted complex and axially n-

propyl substituted complexes are C$_1$ symmetry. The TiFPc and TiCH$_3$Pc are C$_{4v}$ and C$_s$ in symmetry. The silver color central metal ion is titanium, carbon grey, nitrogen blue, hydrogen white, iodine purple, chlorine light green and fluorine dark green. These structures are laid on xy-plane. (For interpretation of the references to color in this figure legend, the reader is referred to the web version of this article.)

## 2.1 Models

Figure 1 depicts the orientation and atom labeling scheme for TiPc molecules. As depicted in Figure 2, nine different TiPc molecules were investigated for their electronic structures and oxygen binding abilities toward both [Ti(IV)Pc]$^{2+}$ and Ti(II)Pc (18 molecules totally) namely, titanyl-phthalocyanine (TiPc), 2,3,2',3'-tetrafluoro-TiPc (TiPc-4F), 2,3,2',3'-tetrachloro-TiPc (TiPc-4Cl), 2,3,2',3'-tetraiodo-TiPc (TiPc-4I) 2,3,2',3'-tetramethyl-TiPc (TiPc-4CH$_3$), and 2,3,2',3'-tetraethyl-TiPc (TiPc-4CH$_2$CH$_3$) as peripherally substituted ligands. In addition, fluoro(phthalocyaninato)titanium (TiFPc), methyl(phthalocyaninato)titanium (TiCH$_3$Pc), and n-propyl(phthalocyaninato)titanium (Ti(CH$_2$)$_2$CH$_3$Pc) were studied as axially substituted ligands to [Ti(IV)Pc]$^{2+}$ and Ti(II)Pc molecules. The halogens were chosen to mimic the electron-withdrawing groups effect on the electronic structure and oxygen reduction, and the methyl, ethyl, and n-propyl molecules were intended to study electron-donating substituents' role in oxygen reduction on [Ti(IV)Pc]$^{2+}$ and Ti(II)Pc phthalocyanines. Methyl, ethyl, and n-propyl groups were helpful in revealing the influence of the size of the substituent on the electron-donating ability of the ligands. The oxygen binding energy, peroxide binding energy, and carbon monoxide binding energies were calculated to investigate the oxygen reduction ability and tolerance to CO poisoning. In addition, the gas-phase structures of Zinc-phthalocyanine (ZnPc) and dichloro(phthalocyaninato)titanium(IV) (TiCl$_2$Pc) molecules were theoretically investigated and compared with the available experimental single crystal X-ray crystallography data to benchmark the theoretical method that has been used to study the novel TiPc molecules.

## 3. Results and discussion

The supplementary materials(SM) accompanying this article provide a comparison of calculated and X-ray adsorption spectroscopy results of Zn(II)Pc (SM Figure S1, Table S1) and TiCl$_2$Pc (SM Figure S2, Table S2) to benchmark the method. The structural trends have been briefly explained in the supplementary materials.

**Table 1** DFT predicted bond distances (Å) and angles (deg) for [Ti(IV)Pc]$^{2+}$ and Ti(II)Pc

|  | Ti(II)Pc | [Ti(IV)Pc]$^{2+}$ |
|---|---|---|
| $M - N_p$ | 2.008 | 1.973 |
| $N_p - C_\alpha$ | 1.396 | 1.407 |
| $C_\alpha - C_\beta$ | 1.452 | 1.441 |
| $C_\beta - C_\beta$ | 1.417 | 1.419 |
| $C_\alpha - N_b$ | 1.331 | 1.322 |
| $C_\beta - C_o$ | 1.399 | 1.398 |
| $C_o - C_m$ | 1.395 | 1.392 |
| $C_m - C_m$ | 1.410 | 1.413 |
| $C_\alpha - N_b - C_\alpha$ | 125 | 125 |
| $C_\alpha - N_p - C_\alpha$ | 109.1 | 108.2 |
| $C_\beta - C_\alpha - N_p$ | 108.2 | 108.5 |
| $C_\beta - C_\beta - C_\alpha$ | 107.2 | 107.4 |
| $C_\beta - C_o - C_m$ | 118.0 | 117.4 |
| $C_o - C_m - C_m$ | 121.1 | 121.4 |

Under the COSMO implicit solvation method with water, the DFT calculations predicted bond lengths and angles for both Ti(II)Pc and [Ti(IV)Pc]$^{2+}$ molecules without any ligand substitution (Table 1). The central atom is chelated by four coordinated-nitrogen atoms in both molecules.

The structure around the central atom with $N_4$ chelated nitrogen is square-planar in shape for $[Ti(IV)Pc]^{2+}$, whereas the square-pyramidal structure can be assigned to Ti(II)Pc. Therefore, the overall symmetry of $D_{4h}$ has been assigned to the $[Ti(IV)Pc]^{2+}$ molecule and $C_{4v}$ to Ti(II)Pc in their optimized ground states. $Ti^{+4}$ ion with +4 charge has smaller ionic radius (0.745 Å) than that of $Ti^{+2}$ ion (1.00 Å) [48], which in principle should make the bond length between the central ion and the $N_p$ nitrogen atom of the $[Ti(IV)Pc]^{2+}$ molecule to be longer than that of Ti(II)Pc. However, due to the $C_{4v}$ symmetry of the ground state of the Ti(II)Pc molecule, the coordinated nitrogen atoms possess relatively longer bond lengths. Thus, the distance of the central metal to $N_p$ nitrogen of Ti(II)Pc was larger than that of $[Ti(IV)Pc]^{2+}$. We also calculated both molecules without COSMO and found that they have $D_{4h}$ symmetry in the gas-phase ground state structure. Thus, it seems that the aqueous environment has greater influence on Ti(II)Pc than on $[Ti(IV)Pc]^{2+}$. These observations inferred that $Ti^{+4}$ binds much stronger to the Pc ligand than $Ti^{+2}$ ion. This aspect of the central ion is further discussed in section 3.2 of this article.

*3.1 Electronic Structure of TiPcs*

Figure 3 shows the molecular orbital (MO) distribution in Ti(II)Pc and $[Ti(IV)Pc]^{2+}$ with the rest of the divalent, uncharged transition metal-phthalocyanine complexes. The Ti(II)Pc molecule is a triplet in the ground state due to the presence of two unpaired electrons in the $1e_g(d_{xz/yz})$ degenerate orbitals, and the $[Ti(IV)Pc]^{2+}$ molecule is singlet in the ground state. Under the $D_{4h}$ point group, five Ti d-orbitals represented as $e_g(d_{xz}, d_{yz})$ degenerate orbitals, $b_{2g}(d_{xy})$, $a_{1g}(d_{z^2})$ and $b_{1g}(d_{x^2-y^2})$. In general, for both titanyl-phthalocyanine (TiPcs) molecules, the $d_{x^2-y^2}$ orbital lies at the highest energy state and the $d_{z^2}$ orbital rests far below from $d_{x^2-y^2}$. The $d_{xy}$ and the degenerate $d_{xz}, d_{yz}$ orbitals lie in energetically lower states. Thus, the orbitals are arranged in the descending order of energy $d_{x^2-y^2} \gg d_{z^2} > d_{xy} > d_{xz}, d_{yz}$. Thus, a square pyramidal crystal field splitting in aqueous environments has been suggested for this class of molecules. The +2 charge of the [Ti(IV)Pc]$^{2+}$ molecule forms coordination complexes with water along the z-axis and thus, may be the reason for high energy $d_{z^2}$ orbital to form a square pyramidal crystal field. This phenomenon was observed in the absorption spectrum for axially substituted [Ti(IV)Pc]$^{2+}$ molecules [39].

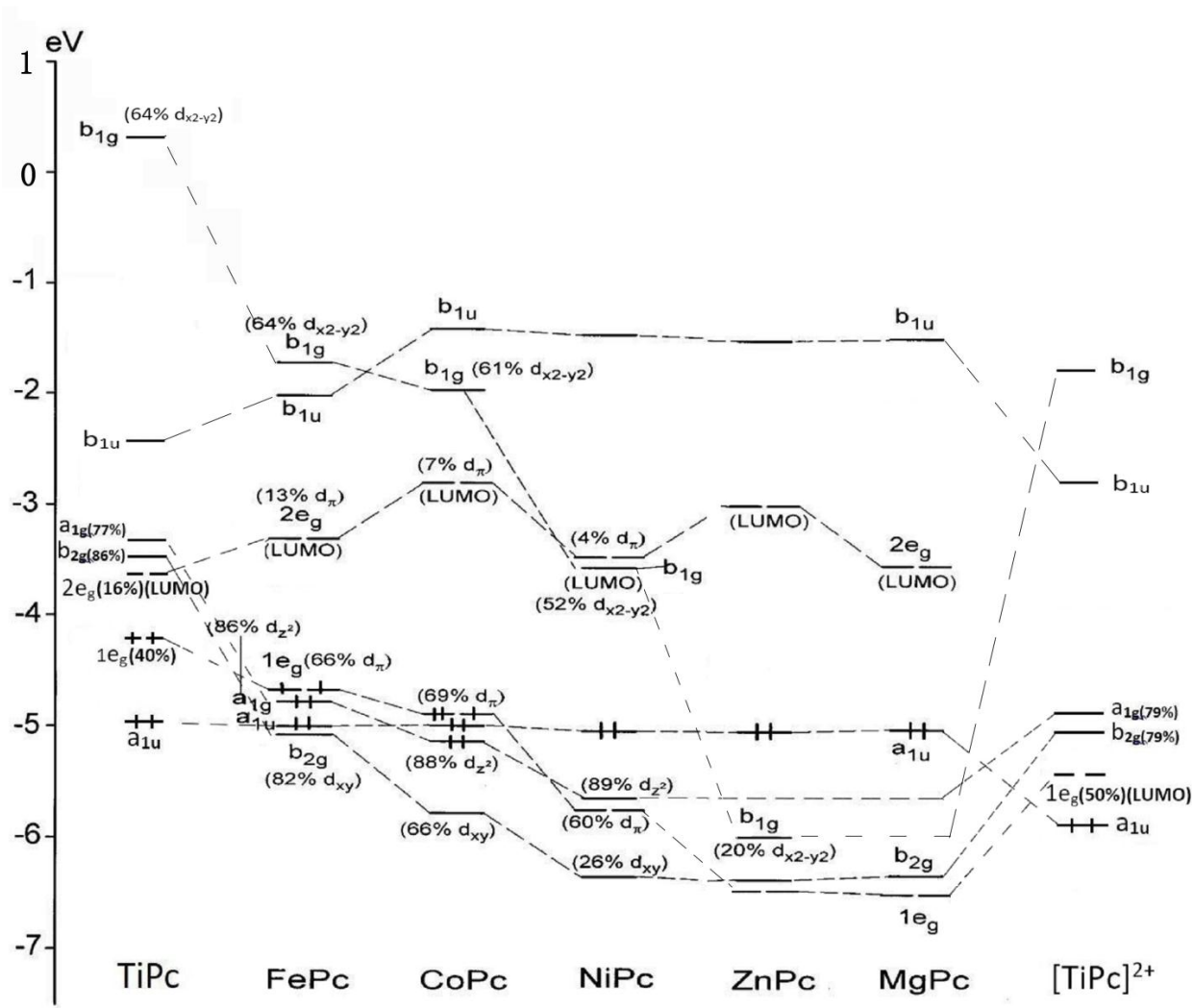

**Figure 3** Electronic structure of Ti(II)Pc and [Ti(IV)Pc]$^{2+}$ with the divalent, uncharged transition metal-phthalocyanine complexes. Except $a_{1u}$ and $b_{1u}$ other low laying Pc ring orbitals are omitted for clarity. Note that D$_{4h}$ point group symmetry notation has been used for the C$_{4v}$ Ti(II)Pc molecule to avoid confusion. Adopted from reference [28].

Due to the triplet state of the Ti$^{+2}$ ion in Ti(II)Pc, the spin-up and spin-down electronic orbitals of $d_{yz}, d_{xz}$ comprised the highest occupied molecular orbital (HOMO) and lowest unoccupied molecular orbital (LUMO), respectively. On the other hand, the [Ti(IV)Pc]$^{2+}$ HOMO consisted of mainly phthalocyanine (Pc) ring $a_{1u}(p_z)$ orbital and the LUMO consisted of empty $1e_g(d_{xz}, d_{yz})$ orbitals. Based on the DFT calculated results, Ti(II)Pc has an $(a_{1u})^2(1e_g)^2$ ground state electronic configuration. Thus, the HOMO lies on spin-up $d_{xz/yz}$ degenerate orbitals, and LUMO lies on the spin-down $d_{xz/yz}$. [Ti(IV)Pc]$^{2+}$ has $(a_{1u})^2(1e_g)^0$ ground state configuration and thus, has two electrons less than the Ti(II)Pc molecule. The percentage values (in parentheses) represent the d-character of some orbitals. It is interesting to note that the d-character of LUMOs increases from right to left and decreases for HOMOs.

*3.1.1 Peripheral and axial ligand substituted complexes of Ti(II)Pc*

The peripherally substituted molecules have spin-up $d_{xz}, d_{yz}$ orbitals as their HOMOs. The HOMOs energies for unsubstituted and peripherally substituted complexes were approximately −4.2 eV. Comparatively, the axially substituted species HOMOs increased in energy, rising to −3.7 eV. The LUMOs are energetically ranged between −3.8 eV to −3.4 eV for peripheral and axial substitutions, respectively.

In general, frontier orbitals in titanyl-phthalocyanines highly hybridized between central metal d-orbitals and phthalocyanine carbon, and nitrogen orbitals. Primarily, $p_z$ phthalocyanine inner ring pyrrole carbon orbitals were observed as LUMOs in halogen complexes. However, it is worth mentioning that the $d_{xz}, d_{yz}$ orbitals were also present in the LUMO with slightly lower percentage than the $p_z$ orbitals. This indicates that the electron-withdrawing effect stabilized the phthalocyanine inner ring carbon orbitals.

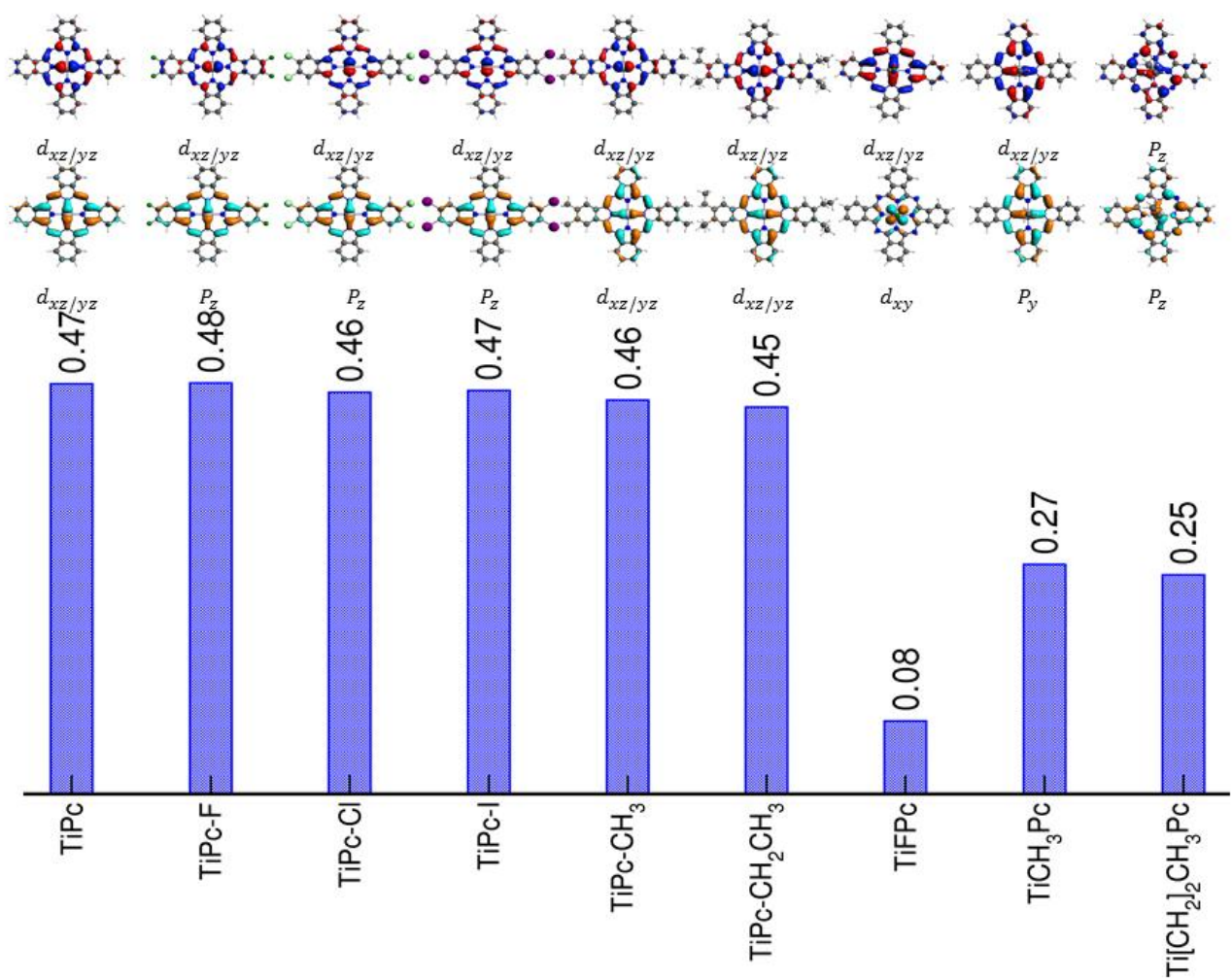

(a)

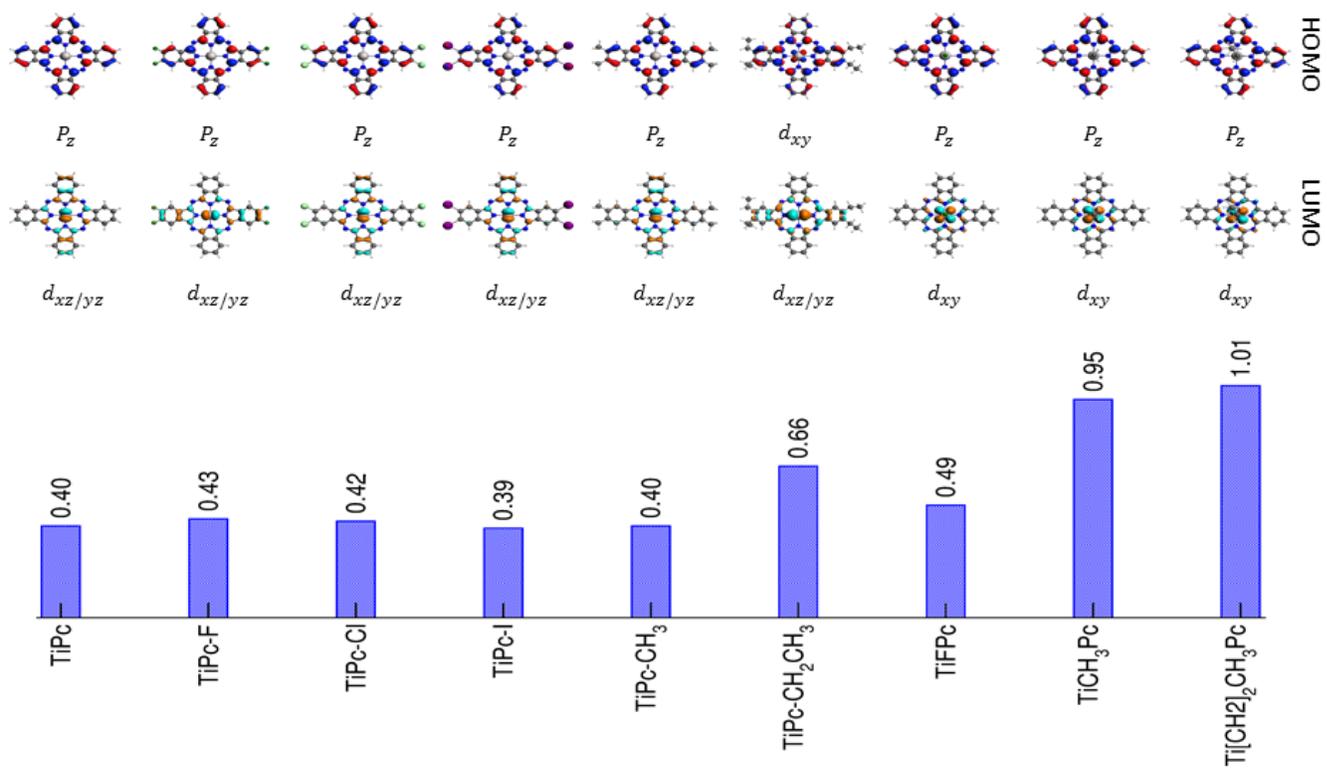

**(b)**

**Figure 4** HOMO-LUMO orbitals and gap energies of (a) triplet state Ti(II)Pc molecules and (b) singlet state [Ti(IV)Pc]$^{2+}$ molecules. HOMO orbitals are blue and red, whereas LUMO orbitals

are light blue and light brown in color. The color of the wave function of the orbitals change according to its sign. Singlet molecules have a tendency to increase GAP energies up to 1 eV from axial ligand substitution, whereas triplet molecules decrease their gap energy due to axial substitution. (For interpretation of the references to color in this figure legend, the reader is referred to the web version of this article.)

The HOMOs and LUMOs of the peripherally substituted methyl and ethyl ligand complexes are very similar to the unsubstituted Ti(II)Pc molecule. It was observed that the HOMO-LUMO gap energies were 0.01–0.02 eV lower than those of the halogen complexes. The larger electron-donating ligand was slightly lower in gap energy in comparison to the smaller electron-donating ligand. This suggests that the larger electron-donating ligand substitution might decrease the gap energy. However, regardless of their electron-donating or withdrawing ability, peripheral ligands do not change the HOMO-LUMO gap significantly.

Among all the ligand substituted complexes, the axially substituted fluorine complex has the lowest gap energy, which is 0.084 eV. In comparison to the unsubstituted molecule and other peripherally substituted complexes gap energies, this molecule should be highly unstable or meta-stable complex. The HOMO is primarily composed of the carbon $d_{xz/yz}$ orbital, and the LUMO consists of the $d_{xy}$ orbital of this molecule. Axially substituted electron-donating complexes have changed the frontier orbitals. Depending on the size of the electron-donating axial substituents, the Pc ring orbitals are stabilized and replace Ti d-orbitals from the frontier orbital positions. The bulk nature of the axial substituent seems to decrease the gap energy, as well. Therefore, relatively larger electron donating-axial substitutions may destabilize the molecule. Nonetheless, all the axial substituted complexes have much lower gap energies. Thus, they can be considered as meta-stable complexes. Figure 4a illustrates all the frontier orbitals and gap energies for Ti(II)Pc peripherally and axially substituted complexes.

The electronic structure of Ti(II)Pc indicates that the central metal is redox active due to both a HOMO and a LUMO that are primarily located in the metal ion d-orbitals in the aqueous environment. In peripherally halogen substituted complexes, metal oxidation and Pc ring reduction can be expected. Moreover, large electron-donating ligand axially substituted complexes have a redox active Pc ring. Thus, the ligand substitutions significantly change the redox processes of the Ti(II)Pc. Peripheral substitution to axial substitution changed the electronic structure of Ti(II)Pc into a redox active central metal ion to a redox active Pc ring molecular complex.

*3.1.2 Peripherally and axially substituted complexes of [Ti(IV)Pc]$^{2+}$*

The point group symmetry of the singlet complexes is similar to the triplet complexes. Contrary to Ti(II)Pc, all of the axial substituted [Ti(IV)Pc]$^{2+}$ molecular complexes are much more stable than their peripherally substituted complexes and the unsubstituted molecules (Figure 4b). In fact, the n-propyl substituted complex has a greater than 1 eV HOMO-LUMO gap energy, and it is the highest recorded for all the molecular systems studied in this article. The HOMO energies were around −5.6 eV to −5.4 eV range for peripherally and axially substituted complexes, respectively. The LUMO energies were observed between −5.2 eV to −4.4 eV energy levels for peripheral and axial substitutions.

Peripherally substituted electron-donating ligands have shown the same trend as axial ligation complexes. Therefore, it is safe to say that the gap energy increases with larger electron-donating group substitutions. Thus, larger electron-donating ligands stabilize the molecular complexes in [Ti(IV)Pc]$^{2+}$.

With the increase in the atomic number of the halogen, the gap energy decreases accordingly in [Ti(IV)Pc]$^{2+}$ complexes. As depicted in Figure 4b, the calculated gap energies were fluorine 0.43 eV, chlorine 0.42 eV, and iodine 0.39 eV. The iodine substituted complex has the lowest energy gap among all the halogen substitutions. The halogen-substituted complexes, in general, have energy gaps comparable to methyl substituted complex. In contrast to Ti(II)Pc complexes, all the HOMOs were composed of phthalocyanine $p_z$ orbitals, except for the ethyl substituted complex. The LUMO was composed of $d_{xz}, d_{yz}$ orbitals for peripherally substituted complexes and $d_{xy}$ for axially substituted complexes. In addition, from the electronic structure of the [Ti(IV)Pc]$^{2+}$ complexes, mainly Pc ring oxidation and central transition metal reduction can be expected. However, it is well known that the highly polarizable metal ions in MPc complexes weaken the ring oxidation process [49]. Therefore, we suspect that [Ti(IV)Pc]$^{2+}$ and their tailored complexes may have weak oxygen reduction ability.

*3.2 Oxygen adsorption*

The chemisorption of $O_2$ on macrocyclic complexes is a result of the interaction between anti-bonding orbitals ($\pi^*$) of the dioxygen molecule and the non-bonding molecular d-orbitals of the macrocyclic complex [31]. Therefore, in principle, this interaction strength can be measured from dioxygen adsorption energies. The bond distance between central metal on the macrocyclic complex and adsorbed dioxygen can be used as an indirect measurement of adsorption strength, as well.

According to literature, there are four possible modes of oxygen binding to transition metal macrocyclic complexes [50,51], namely, side-on, end-on, bridge, and trans modes. Among them, the side-on and end-on modes are the primary binding modes of dioxygen to MPcs. However, in this study, we have focused only on the side-on binding of a dioxygen molecule to titanyl-phthalocyanines. This is because the systems we studied were unable to form stable end-on adducts with a dioxygen molecule. Therefore, the side-on adducts are the most stable form of dioxygen adducts in TiPcs. The optimized dioxygen-adduct structures are illustrated in Figure S3 of supporting materials. Their O-O bond distances ($R_{O-O}$), metal to oxygen distances ($R_{M-O}$), AIM charges of titanium and averaged charge of adsorbed dioxygen atom, and oxygen adsorption energies are listed in Table 2. Note that all the molecular complexes are denoted by similar symbols for simplicity, as in Figure 1. The oxygen adsorption energies for TiPcs were evaluated from equation-1,

$$E_{b_{O2}} = E_{ML-O_2} - [E_{O2} + E_{ML}] \quad (1)$$

where $E_{ML-O_2}$ is the total energy of the molecule-oxygen adduct, $E_{O2}$ is the energy of the isolated oxygen molecule, and $E_{ML}$ is the energy of the titanyl-phthalocyanine complex. The modified version of this equation has been used for hydrogen peroxide and carbon monoxide adsorption energy calculations.

The dioxygen adsorption energy results indicated that the Ti(II)Pc complex and its derivatives form more stable dioxygen-adducts than the corresponding [Ti(IV)Pc]$^{2+}$ complex and its derivatives. It is very clear from our results that electron-donating substituents facilitate the oxygen adsorption process. Thus, the dioxygen adsorption energies are increased in Ti(II)Pc electron-donor substituents complexes (with respect to unsubstituted Ti(II)Pc). In fact, the highest oxygen adsorption energy was related to the largest electron-donating peripheral ligand substituent. The electron-withdrawing substituents decreased the electron donating ability of the central metal to the oxygen molecule, consequently decreasing the oxygen adsorption energy.

However, the electron-donating and electron-withdrawing effects are not very clear in [Ti(IV)Pc]$^{2+}$ complexes. Although iodine is an electron-withdrawing substituent, it increases the oxygen adsorption energy. In contrast, ethyl is an electron-donating substituent, but it decreases the oxygen adsorption energy. This can be rationalized by the DFT calculated gap energies. Among the peripherally substituted electron-withdrawing ligands (halogens) in [Ti(IV)Pc]$^{2+}$, iodine has the lowest gap energy; therefore, it is the least stable halogen complex. Thus, it naturally tends to react to form stable complexes and facilitates electron transfer to the oxygen molecule, forming more stable dioxygen adduct with higher oxygen adsorption energy. To confirm this hypothesis, we calculated AIM (Bader) effective atomic charges of oxygen atoms in dioxygen adduct. We found that indeed the [Ti(IV)Pc]$^{2+}$ with an iodine ligand has significant charge transfer towards oxygen molecule (Table 2). On the contrary, the ethyl substituted complex has the highest gap energy; hence, it is already a more stable molecular complex than the other peripheral substituted complexes. Thus, the ethyl-substituted complex is less reactive than the methyl substituted complex. The Bader partial charges also showed a slight reduction of charges on oxygen atoms in the ethyl complex. Consequently, the ethyl substituted [Ti(IV)Pc]$^{2+}$ complex has a lower oxygen adsorption energy.

Nevertheless, axially substituted complexes are the least stable complexes with the lowest oxygen adsorption energies for both TiPc complexes. In the Ti(II)Pc axial derivatives, the fluorine substituted [FTi(II)Pc]$^-$ complex has the lowest oxygen adsorption energy, whereas methyl and n-propyl substituted complexes have slightly higher adsorption energies. Thus, the larger the size of the electron-donating substituent, the more stable the dioxygen adduct complex formed. This situation is similar for both the TiPc axial substituted complexes. In the case of $[Ti(IV)Pc]^{2+}$, axially fluorine substituted $[FTi(IV)Pc]^+$ complex has the lowest oxygen binding energy. The positive sign of the energy indicated that the oxygen molecule does not bind to the complex. With the help of Bader charge analysis and DFT predicted HOMO-LUMO gap energies, we can rationalize non-bonding of the oxygen molecule to $[FTi(IV)Pc]^+$ complex. The highly positive Bader charge on titanium ion in this complex indicated that direct withdrawal of electrons from central ion reduced the electron density. As a result, titanium ion reduced the ability to transfer electrons to the oxygen molecule. Nonetheless, the stability of the molecule determines the reactivity towards oxygen. Therefore, the more stable $[FTi(IV)Pc]^+$ molecule did not react with the oxygen molecule to form dioxygen adduct, whereas the highly unstable $[FTi(II)Pc]^-$ complex reacted with the oxygen molecule to form an adduct. The stability factor also determines the magnitude of the adsorption energy. The highly stable axial ligand complexes in $[Ti(IV)Pc]^{2+}$ have lower adsorption energies than meta-stable Ti(II)Pc axial ligand complexes. In addition, compellingly higher charge transfer from $[H_3CTi(IV)Pc]^+$ and $[H_3C(CH_2)_2Ti(IV)Pc]^+$ complexes to oxygen molecule may be the reason that they formed dioxygen-adduct complexes though they are highly stable molecules.

Moreover, the differences between Ti(II)Pc and $[Ti(IV)Pc]^{2+}$ charge transfer to an oxygen molecule may be attributed to the loosely bound, slightly out-of-the-$N_4$-plane, over-coordinated $Ti^{+2}$ ion and its higher electron density. In fact, the Bader bond critical point calculations of these two complexes revealed that $Ti^{4+}$ strongly binds to coordinated nitrogen atoms with 0.1133 bond critical point density. However, $Ti^{2+}$ has rather weak coordination bonds with a 0.1016 critical point density. The out-of-the-plane position of the $Ti^{+2}$ ion can facilitate the oxygen adsorption reaction in two ways: by reducing the steric hindrance and by reducing the overlap between metal d-orbitals with the Pc ring orbitals. The latter may ease the metal to dioxygen charge transfer process by preventing delocalization of electrons in the Pc ring. Therefore, Ti(II)Pc molecule and its derivative complexes are highly reactive towards molecular oxygen in the aqueous phase. This behavior also explains the higher oxygen adsorption energy of Ti(II)Pc complexes. We also compared atomic oxygen adsorption energy of unsubstituted TiPc molecules with Zr atomic oxygen adsorption energy (~8 eV), which are in good agreement. A study [52] showed an increase in oxygen adsorption energy on transition metals from right to left in the periodic table. Therefore, it is quite expected to have higher dioxygen adsorption energy for titanium.

The $R_{O-O}$ and $R_{M-O}$ values in Table 2 also predicted that the weakest dioxygen adsorption molecules are $[FTi(IV)Pc]^+$ and $[FTi(II)Pc]^-$. These molecules have the shortest O-O bond distances and longest M-O distances, indicating weak adsorption. Correspondingly, the longest O-O bond length and shortest M-O bond length should represent the strongest dioxygen adsorption, as observed in the Ti(II)Pc molecular complexes. As expected from the above results, the peripherally substituted ethyl complex has the longest O-O bond distance and the shortest M-O bond distance. However, in $[Ti(IV)Pc]^{2+}$ complexes, the longest O-O bond length and the shortest M-O bond length are reported in axial electron donating group substituted ligand complexes. This is an effect of higher charge transfer ability of these complexes. Thus, higher electron accumulation in the oxygen molecule tends to elongate the O-O bond.

In addition, we found all the adsorbed dioxygen bond lengths were around ~1.5 Å for Ti(II)Pc complexes, whereas for $[Ti(IV)Pc]^{2+}$ complexes, this value varied from 1.3 Å to 1.4 Å depending on the charge transfer of the complex. The high-charge transferring electron-donating axial ligation and iodine-substituted complex have 1.4 Å O-O bond lengths. According to Cramer et al. [53], 1.2 Å to 1.3 Å adsorbed dioxygen bond lengths are considered as superoxo ($O_2^-$) type species, and the 1.4 Å to 1.5 Å bond lengths are categorized as peroxo ($O_2^{2-}$) type species. In our studies of TiPc complexes, we found both the categories. The primarily adsorbed dioxygen in Ti(II)Pc are the peroxo type, and [Ti(IV)Pc]$^{2+}$ are the superoxo type. Thus, the character of the adsorbed dioxygen species is determined by the charge transfer ability of the parent molecule.

**Table 2** Calculated dioxygen-adduct properties and Bader charges for all of the complexes studied. M is multiplicity $R_{O-O}$ and $R_{M-O}$ are dioxygen bond distance and metal to oxygen distances, respectively. Bond lengths are recorded in Å, and adsorption energy is recorded in eV.

| Compound | M | Ti | O | $R_{O-O}$ | $R_{M-O}$ | $E_bO_2$ |
|---|---|---|---|---|---|---|
| TiPc | 3 | 1.941 | -0.527 | 1.462 | 1.858 | -3.19 |
| TiPc-F | 3 | 1.962 | -0.521 | 1.470 | 1.853 | -3.07 |
| TiPc-Cl | 3 | 1.954 | -0.521 | 1.461 | 1.858 | -3.15 |
| TiPc-I | 3 | 1.953 | -0.521 | 1.466 | 1.852 | -3.08 |
| TiPc-CH$_3$ | 3 | 1.948 | -0.529 | 1.467 | 1.855 | -3.19 |
| TiPc-CH$_2$CH$_3$ | 3 | 1.947 | -0.527 | 1.467 | 1.854 | -3.21 |
| TiFPc | 3 | 2.079 | -0.543 | 1.439 | 1.910 | -1.37 |
| TiCH$_3$Pc | 3 | 1.956 | -0.549 | 1.464 | 1.868 | -1.93 |
| Ti(CH$_2$)$_2$CH$_3$Pc | 3 | 1.948 | -0.551 | 1.464 | 1.869 | -2.18 |
| TiPc | 1 | 2.086 | -0.230 | 1.332 | 1.972 | -2.18 |
| TiPc-F | 1 | 2.113 | -0.230 | 1.332 | 1.968 | -2.12 |
| TiPc-Cl | 1 | 2.091 | -0.232 | 1.333 | 1.968 | -2.14 |
| TiPc-I | 1 | 2.084 | -0.483 | 1.457 | 1.850 | -2.36 |
| TiPc-CH$_3$ | 1 | 2.085 | -0.263 | 1.344 | 1.957 | -2.17 |
| TiPc-CH$_2$CH$_3$ | 1 | 2.073 | -0.261 | 1.344 | 1.958 | -1.92 |
| TiFPc | 1 | 2.129 | -0.247 | 1.312 | 2.076 | 0.20 |
| TiCH$_3$Pc | 1 | 2.012 | -0.494 | 1.458 | 1.848 | -0.86 |
| Ti(CH$_2$)$_2$CH$_3$Pc | 1 | 1.991 | -0.497 | 1.458 | 1.849 | -1.05 |

*3.3 $H_2O_2$ adsorption*

$$XM^{II} + O_2 \Leftrightarrow XM^{\delta+} \ldots O_2^{-\delta} \quad (2)$$

$$XM^{\delta+} \ldots O_2^{\delta-} + H^+ \rightarrow (XM^{III} \ldots O_2H)^+ \quad (3)$$

$$(XM^{III} \ldots O_2H)^+ + H^+ + 2e^- \rightarrow XM + \text{intermediates } (H_2O_2) \quad (4)$$

$$H_2O_2 + XM \rightarrow H_2O + \tfrac{1}{2} O_2 \quad (5)$$

Beck [48] and Zagal et al. [54] proposed the mechanism of oxygen reduction in phthalocyanine and porphyrin macrocyclic complexes. According to them, the first step of oxygen reduction is the dioxygen adsorption on the macrocyclic complex (Eq 2). Then, the next step is the electron transfer from the transition metal center of the complex to the adsorbed oxygen molecule that forms intermediate complexes (Eqs 3 and 4). This intermediate complex in acidic media is a hydrogen peroxide molecule. Finally, the O-O bond of the peroxide molecule breaks on the active site of the catalysts to form two OH molecules, and they interact with hydronium ions in the surrounding environment to produce water (Eq 5).

Therefore, the most important reaction step that determines the complete reduction of the oxygen molecule into water is the O-O bond splitting of the peroxide molecule. If the macrocyclic complex breaks the O-O bond of the peroxide, it will lead to water formation via a 4e$^-$ oxygen reduction mechanism. If the molecule is unable to break the peroxide bond, the oxygen reduction will be limited to the formation of peroxide. This was considered a 2e$^-$ oxygen reduction pathway.

**Table 3** $H_2O_2$ adsorption energies and calculated properties of adsorbed complexes. M is the multiplicity of the complexes; $R_{O-O}$ and $R_{M-O}$ are peroxide oxygen bond distances and metal to oxygen distances, respectively. Bond lengths are recorded in Å, and adsorption energies are recorded in eV.

| Compound | M | $R_{O-O}$ | $R_{M-O}$ | $E_{bH2O2}$ |
|---|---|---|---|---|
| TiPc | 3 | 2.578 | 1.891 | -4.31 |
| TiPc-F | 3 | 2.430 | 1.909 | -4.28 |
| TiPc-Cl | 3 | 3.026 | 1.825 | -3.99 |
| TiPc-I | 3 | 2.250 | 1.897 | -0.87 |
| TiPc-CH$_3$ | 3 | 2.579 | 1.893 | -4.38 |
| TiPc-CH$_2$CH$_3$ | 3 | 2.461 | 1.883 | -4.34 |
| TiFPc | 3 | 1.474 | 2.497 | -0.42 |
| TiCH$_3$Pc | 3 | 1.471 | 2.628 | -0.33 |
| Ti(CH$_2$)$_2$CH$_3$Pc | 3 | 1.473 | 2.666 | -0.32 |
| TiPc | 1 | 1.467 | 2.135 | -1.53 |
| TiPc-F | 1 | 1.472 | 1.853 | -1.57 |
| TiPc-Cl | 1 | 2.430 | 1.853 | -4.01 |
| TiPc-I | 1 | 1.464 | 2.131 | -1.52 |
| TiPc-CH$_3$ | 1 | 2.102 | 1.892 | -1.49 |
| TiPc-CH$_2$CH$_3$ | 1 | 1.465 | 2.134 | -1.22 |
| TiFPc | 1 | 1.465 | 2.580 | -0.24 |
| TiCH$_3$Pc | 1 | 1.466 | 2.762 | -0.18 |
| Ti(CH$_2$)$_2$CH$_3$Pc | 1 | – | – | – |

To understand the capability of these systems for complete reduction of $O_2$ to water, all of the molecules under this study were optimized with the $H_2O_2$ molecule using a COSMO water solvent environment. The DFT-calculated properties and $H_2O_2$ adsorption energies of the optimized structures can be found in Table–3. It has been observed that the axially n-propyl substituted singlet complex did not adsorb peroxide at all. Thus, this adsorption energy is not recorded in the table. Unsubstituted Ti(II)Pc and all of the peripherally substituted species, regardless of the electron-donating or electron-withdrawing groups, spontaneously split the peroxide O-O bond to form two OH groups on the central metal ion as an active site. This is considered a barrier-less dissociative adsorption of peroxide molecules. Therefore, this class of complexes is capable of the complete reduction of $O_2$ to water. Platinum is capable of nearly spontaneous peroxide reduction, but there is still a small (0.22 eV) thermally activate barrier to it [55]. Axially substituted Ti(II)Pc molecules are unable to break the peroxide bonds spontaneously. This is further confirmed by the $[Ti(IV)Pc]^{2+}$ axial substitutions. As we expected from the electronic structure, none of the $[Ti(IV)Pc]^{2+}$ substituted molecules, except for the peripherally chlorine substituted complex, are capable of breaking peroxide O-O bond spontaneously. However, the peripherally chlorine substituted complex can undergo spontaneous reduction of peroxide to hydroxyl. In addition to chlorine, two other peripherally substituted halogen ligands could not undergo spontaneous reduction.

Although the mechanism of this spontaneous dissociative adsorption is not clear, we can theorize the possibility using the calculated AIM charges on $Ti^{4+}$ ion and oxygen atoms of the $H_2O_2$ molecules. Table 4 lists the Bader charges on $Ti^{4+}$ ion and the adsorbed oxygen atoms of the $H_2O_2$ molecules. According to the calculated values, iodine has the least influence on the central metal ion in comparison to the parent $[Ti(IV)Pc]^{2+}$ molecule. This can be attributed to the long distance (2.097 Å) of iodine atoms from the covalently bonded Pc aza benzyl ring carbon. On the other hand, chlorine significantly influenced the $Ti^{4+}$ ion and increased the charge transfer to oxygen atoms in the peroxide molecules. The highly positive effective charge (2.154) on the central metal ion and highly negative effective charges (-1.073, -1.084) on oxygen atoms are evidence for chlorine facilitated charge transfer effects. Therefore, the chlorine-substituted complex [Ti(IV)PcCl]$^{+2}$ is capable of spontaneous dissociative adsorption of $H_2O_2$. In addition, the distance from the chlorine to the aza benzyl carbon (1.727 Å) is shorter than that of iodine. This may be a decisive factor for the strong influence of chlorine. Although the fluorine to the aza benzyl carbon distance is 1.351 Å, it has not significantly influenced the charge transfer effect. This may be attributed to the high electronegativity of the fluorine. The highly electronegative fluorine attracts electrons of the central ion toward itself, delocalizing the Ti$^{4+}$ electrons in the Pc ring. Thus, it reduced the electron donating ability of the central ion to adsorbate molecule. As a result, it does not facilitate charge transfer as effectively as in the peripherally chlorine substituted complex.

**Table 4** The AIM (Bader) charges on titanium and peroxide oxygen atoms of parent [Ti(IV)Pc]$^{2+}$, Iodine, Chlorine and Fluorine peripherally substituted complexes. Distance between peripherally substituted halogen atom to its covalently bonded aza benzyl carbon is also reported.

| Compound | Bader Charges | | | Distance (Å) |
|---|---|---|---|---|
| | Ti$^{4+}$ | O$_1$ | O$_2$ | |
| [Ti(IV)Pc]$^{2+}$ | 2.051 | -0.536 | -0.545 | |
| [Ti(IV)Pc-I]$^{2+}$ | 2.057 | -0.536 | -0.543 | 2.097 |
| [Ti(IV)Pc-Cl]$^{2+}$ | 2.154 | -1.073 | -1.084 | 1.727 |
| [Ti(IV)Pc-F]$^{2+}$ | 2.041 | -0.436 | -0.603 | 1.351 |

Nevertheless, we found a reasonable correlation between the hydrogen peroxide adsorption energies and the spontaneous peroxide O-O bond breaking. According to the calculated peroxide adsorption energies in Table 3, the highest adsorption energy is recorded for the peripherally chlorine substituted singlet complex. It corresponds to the high charge transfer ability of the molecule discussed above. This value for peroxide adsorption in the chlorine complex is approximately 2.5 eV higher than that of other peripheral substitutions.

This also correlates well with Ti(II)Pc peripheral substitutions, except for iodine. We suspect that the higher electron transfer ability in the Ti(II)Pc iodine substituted complex enabled the peroxide bond break.

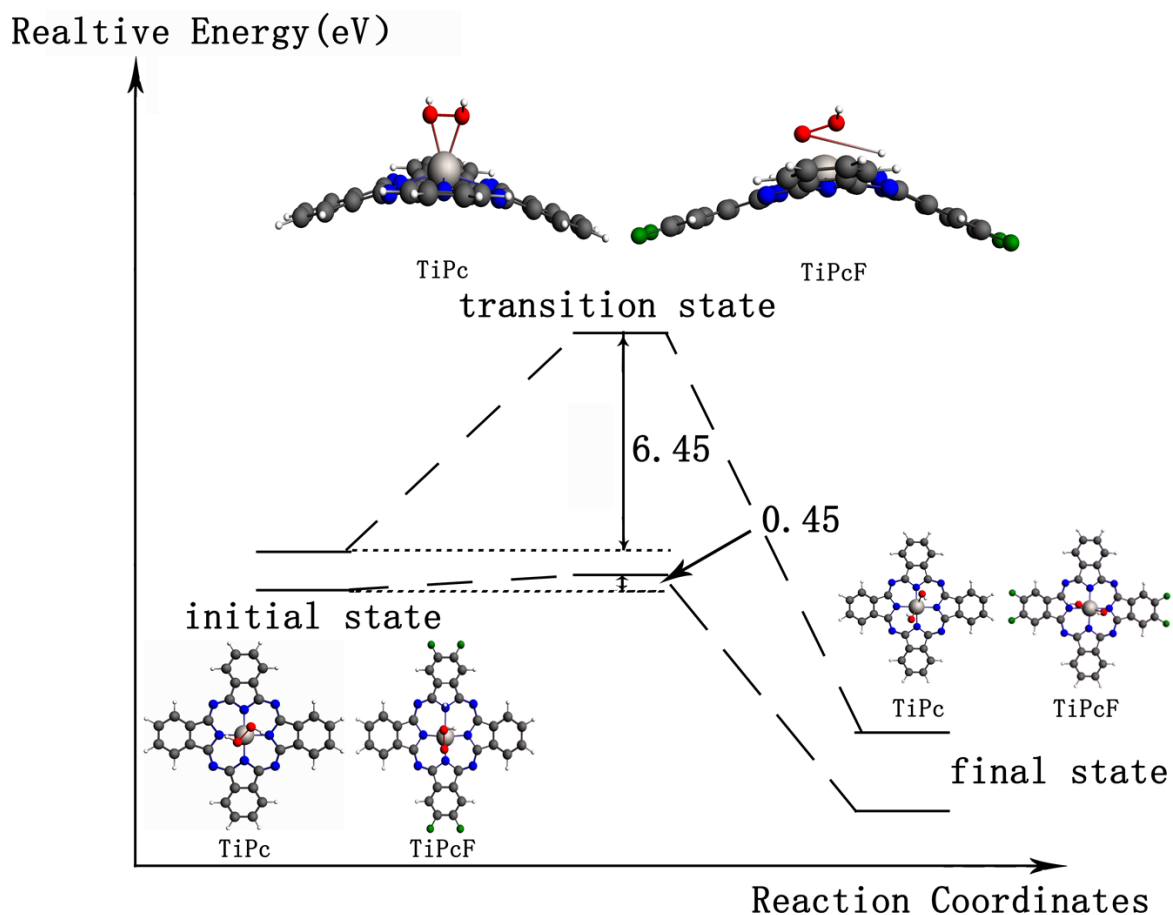

**Figure 5** Relative energy vs reaction coordinates diagram. The activation energy for [$H_2O_2$TiPc]$^{2+}$ singlet complex is 6.45 eV. Peripheral Fluorine substitution to parent singlet complex reduced the activation energy barrier upto 0.45 eV. Initial reactants, transition state and As seen from the results of this study, the peroxide adsorption energies are more fundamental in determining the oxygen reduction ability of the catalyst complexes. Thus, we can claim that the higher the peroxide adsorption energy, the better is the oxygen reduction ability of the titanyl-phthalocyanine complexes.

Furthermore, we calculated the activation energy barrier for the parent $[Ti(IV)Pc]^{2+}$ complex and found it to be 6.45 eV. The initial reactant molecular structures, transition state structures and final product structures are illustrated in Figure 5. The substantial reduction of the activation energy by peripheral fluorine substitution of $[Ti(IV)Pc]^{2+}$ catalysts has been graphically illustrated. The energy barrier is reduced to 0.45 eV, and it can be thermally activated under the PEM fuel cell operating conditions. Axially ligand substituted singlet and triplet complexes were not studied for energy barriers due to their end-on mode of peroxide adsorption (see the supporting information Figure S4).

The intermediate complexes of the oxygen reduction process were also investigated. Figure 6 illustrates the four main intermediate complexes namely, $O_2$-adsorbed complex, OH-adsorbed complex, $H_2O$-adsorbed complex, and $H_2O_2$-adsorbed complexes. It is noteworthy that $H_2O_2$ adsorbed unsubstituted complexes are different in the singlet and triplet spin states. In the triplet spin state, barrier-less dissociative adsorption of peroxide produced two hydroxyl groups spontaneously. In addition, Tables-S3 and S4 listed the bonding energies and adsorption energies of OH and $H_2O$, respectively. The triplets have much lower OH and $H_2O$ adsorption energies than their $O_2$ and $H_2O_2$ adsorption energies. This is an indication of good catalysts for ORR. Due to the weak dioxygen and $H_2O_2$ adsorption on axially substituted complexes (see Tables-2 and 3), they were not investigated for OH and $H_2O$ adsorption.

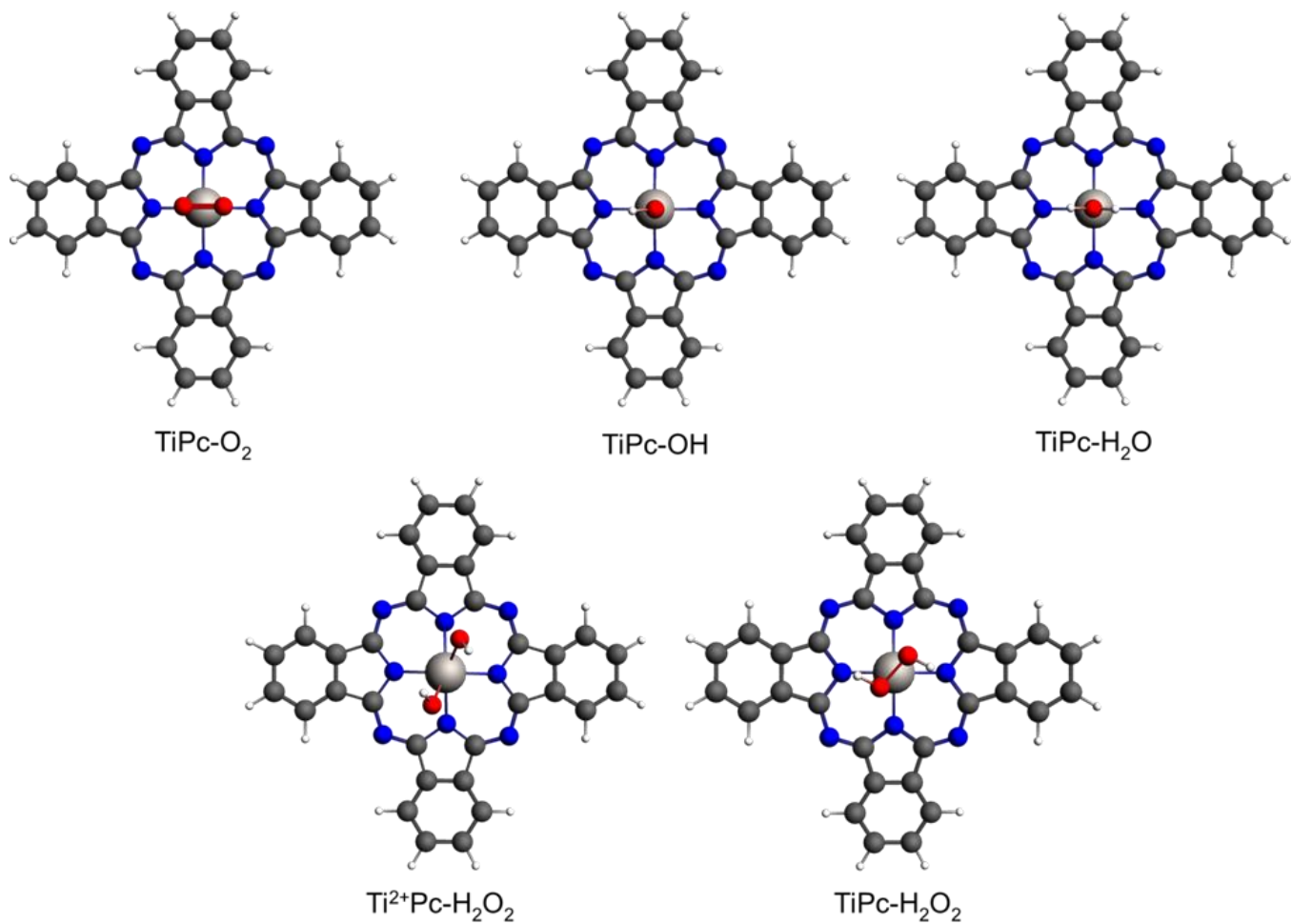

**Figure 6** Intermediates of oxygen reduction reaction and their binding modes in unsubstituted TiPcs. Note that in triplet state parent molecule spontaneously reduced peroxide into hydroxyl. Here carbon grey, oxygen red, nitrogen blue, titanium silver and hydrogen is white in color. (For interpretation of the references to color in this figure legend, the reader is referred to the web version of this article.)

*3.4 CO adsorption*

**Table 5** Representative cases of CO adsorption on singlet and triplet molecules. M is the multiplicity of the complexes. Their CO adsorption energies and AIM effective charges on Ti, C and O are listed. Ti-C, Ti-O and C-O distances of the adsorbed complexes are also reported in Ångstroms.

| Compound | M | Ti | O | C | Ti-O (Å) | Ti-C (Å) | C-O (Å) | $E_{CO}$ (eV) |
|---|---|---|---|---|---|---|---|---|
| COTiPc | 3 | 1.87 | -1.12 | 0.78 | 2.30 | 2.24 | 1.18 | -0.18 |
| COTiPc-Cl | 3 | 1.88 | -1.12 | 0.79 | 2.29 | 2.25 | 1.18 | -0.15 |
| COTiPc-$CH_3$ | 3 | 1.86 | -1.10 | 0.75 | 2.45 | 2.23 | 1.17 | -0.18 |
| COTiPc | 1 | 1.96 | -0.97 | 1.09 | - | 2.23 | 1.13 | -1.00 |
| COTiPc-Cl | 1 | 2.03 | -1.17 | 1.23 | 2.12 | - | 1.15 | -0.36 |
| COTiPc-$CH_3$ | 1 | 2.03 | -1.17 | 1.23 | 2.11 | - | 1.15 | -0.37 |

Carbon monoxide (CO) poisoning in state-of-the-art PEMFCs is one of the major drawbacks in platinum (Pt) catalysts. Thus, Pt has often been alloyed with ruthenium (Ru) in contemporary fuel cells to compensate for its poisonous effect. The proposed novel catalyst system can be utilized for hydrogen oxidation as well. Particularly, phthalocyanine can easily reduce $H_2$ addition to the titanium ion. We investigated CO adsorption on all the peripherally substituted triplet Ti(II)Pc catalysts including unsubstituted Ti(II)Pc. Chlorine and methyl substituted catalysts from [Ti(IV)Pc]$^{2+}$ have been studied to account for electron-withdrawing and electron-donating ligand effects, as well as an unsubstituted singlet complex. Figure 7 illustrates optimized structures of the molecular complexes with adsorbed CO. The calculated bond distances, AIM atomic charges on Ti, O, C, and CO adsorption energies of the complexes are listed in Table 5.

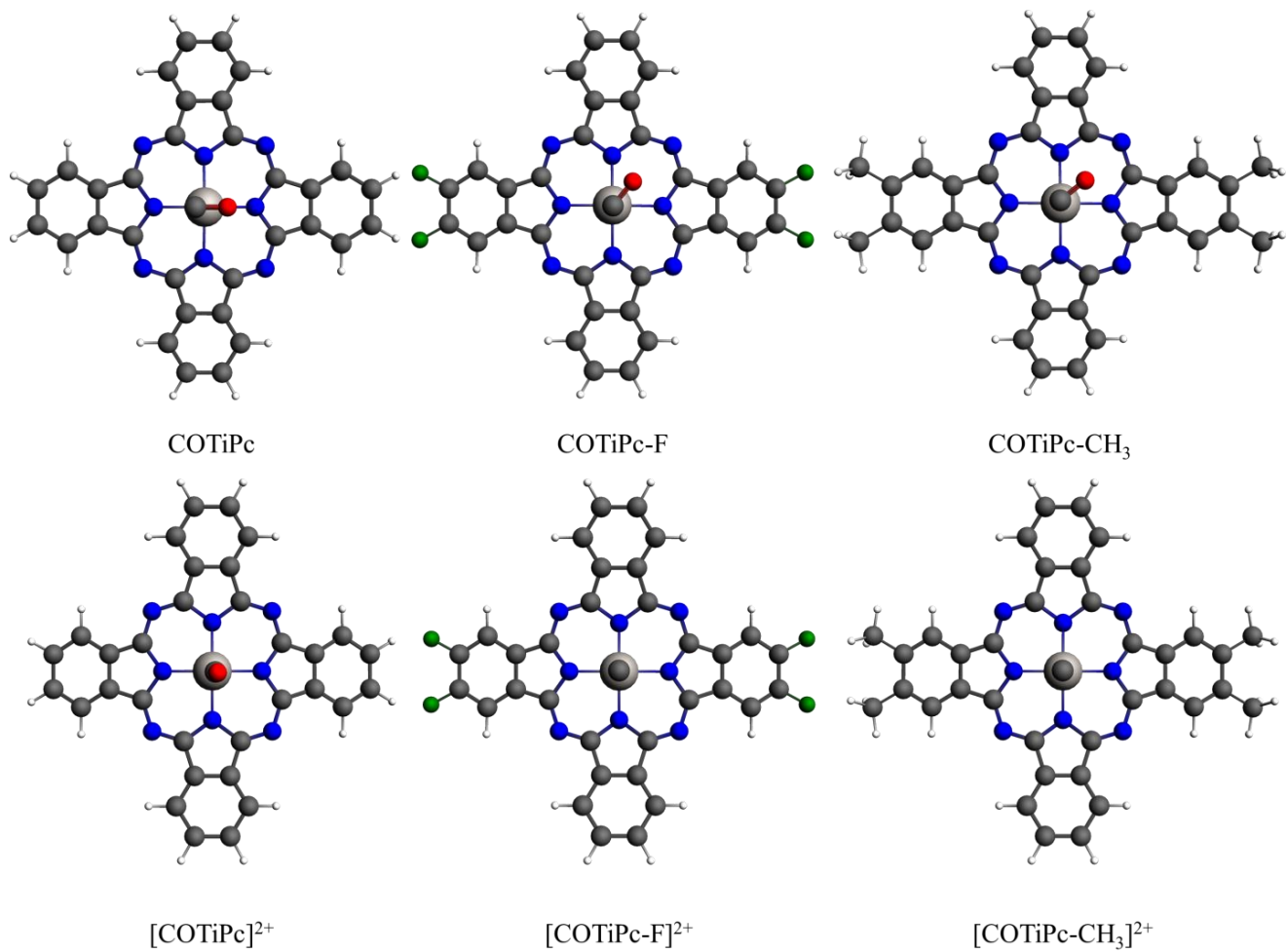

**Figure 7** Carbon monoxide (CO) adsorption on singlet and triplet states of the molecular complexes. Triplet complexes have side-on adsorption, whereas singlet complexes have end-on adsorption. Here carbon grey, oxygen red, nitrogen blue, titanium silver, fluorine dark green and hydrogen is white in color. (For interpretation of the references to color in this figure legend, the reader is referred to the web version of this article.)

There is a noticeable difference in the adsorption modes of the Ti(II)Pc and [Ti(IV)Pc]$^{2+}$ complexes. The CO molecule has end-on adsorption in [Ti(IV)Pc]$^{2+}$ complexes, whereas side-on adsorption is preferred in Ti(II)Pc complexes. In addition, peripherally substituted singlets directly bind to the oxygen atom in CO, whereas in other complexes, Ti directly binds to the carbon atom in CO. This difference can be reconciled with the help of AIM charge analysis on Ti, C and O atoms. Based on the population analysis of partial charges distributed on atoms, singlet complexes have higher positive charge accumulation on the Ti ion and carbon, and higher negative charge accumulation on the oxygen atom. Due to the higher positive charges on Ti and carbon in peripherally substituted singlets, Ti and C strongly repelled from each other. Thus, Ti directly binds only to oxygen in these two peripherally substituted singlet molecules. The slightly higher negative charge on oxygen and higher positive charge accumulation on Ti made the CO adsorption energy twice as strong as in peripherally substituted singlets in comparison to triplets. As a result, CO has a relatively higher adsorption on singlet complexes. Moreover, Ti-O bond distances in singlets are shorter than that of triplets, which provide further evidence for stronger CO adsorption. However, the highest adsorption energy (-1.00 eV) for singlets is reported for unsubstituted complex. In contrast to peripherally substituted complexes, Ti ion directly binds to the carbon atom of CO molecule in [Ti(IV)Pc]$^{+2}$. Thus, the amount of positive charge accumulation on Ti and C atoms determines the atom of CO molecule that binds to singlet TiPcs during the adsorption process. Nevertheless, in singlet complexes, peripheral ligand substitution reduces the CO adsorption energy dramatically. According to Table 5, triplet complexes have the lowest CO adsorption energies. Therefore, triplet complexes can be considered as potential anode catalysts for PEMFCs.

## 4. Conclusions

In this article, we investigated Ti(II)Pc with a triplet spin state and [Ti(IV)Pc]$^{2+}$ with a singlet spin state using a COSMO water implicit solvation model for the first time. The methodology was benchmarked using experimentally and theoretically well-characterized ZnPc and TiCl$_2$Pc complexes. The singlet and triplet titanyl-phthalocyanine molecules were tailored with weak electron-donating and electron-withdrawing ligands in peripheral positions and axial positions of the complexes. The electronic structure of the unsubstituted molecules was compared to the existing DFT results of other transition metal phthalocyanines. Our model electronic structures agreed well with the trends in 3d-transition metal-phthalocyanine complexes.

The DFT calculations predicted that the peroxide binding energies are more relevant to correlate TiPcs. Thus, higher peroxide binding energy resulted in better catalyst performance. The axial ligand groups are considered as the most influential ligands to the central transition metal atom. However, in this study, we theoretically provided evidence for significant peripheral ligand effects on singlet central metal ion. Peripherally substituted halogens dramatically reduced the activation energy barrier of the [Ti(IV)Pc]$^{2+}$ complex from 6.45 eV to 0.45 eV. This is further reduced by chlorine substitution to achieve spontaneous peroxide reduction.

The calculations revealed that the unsubstituted and peripherally substituted Ti(II)Pcs are the best performers in oxygen reduction. This class of molecules has higher dioxygen and peroxide binding energies and significantly lower OH and H$_2$O binding energies. In addition, lower CO binding energies prevented CO poisoning of the Ti(II)Pc catalysts. Moreover, the spontaneous reduction of peroxide O-O bond to proceed 4e$^-$ oxygen reduction can be considered as an extraordinary catalytic ability of triplet Ti(II)Pc catalysts. These properties directly correlated with the DFT calculated electronic structures of the complexes. Therefore, the spontaneous peroxide reduction of the triplet state complexes and the chlorine substituted singlet complex, their high hydrogen peroxide adsorption, and their low CO adsorption made them excellent potential candidates for the PEMFC catalyst.

**Acknowledgment**

JRD thanks the Chinese Scholarship Council (CSC) for PhD scholarship. He also acknowledged the Jacobs-University of Bremen for the computation time provided for this work and the Thomas Heine group for fruitful discussions to improve the manuscript. SZ thanks the AVL GmbH for continuous financial support to his research work. Authors acknowledged the Elsevier language editing service for editing this manuscript to eliminate phrasing issues and typos.

**Competing Financial interests**

The authors declare no competing financial interests.

**Author Contributions**

J.R.D conducted the DFT study and wrote the manuscript. T.H. analyzed the data. S.Z. did the overall supervision of the research and commented on the manuscript.